\newcommand{\PHCX} {(C$_4$H$_{12}$N$_2$)Cu$_2$(Cl$_{1-x}$Br$_{x}$)$_6$ \xspace}

\newcommand{\IPACX}{(CH$_3$)$_2$CHNH$_3$Cu(Cl$_{1-x}$Br$_x$)$_3$ \xspace}
\newcommand{\DTN} {NiCl$_2$$\cdot$4SC(NH$_2$)$_2$\xspace}
\newcommand{\DTNX} {Ni(Cl$_{1-x}$Br$_x$)$_2$$\cdot$4SC(NH$_2$)$_2$ \xspace}

\newcommand{\TCX} {Tl$_{1-x}$K$_x$CuCl$_3$\xspace}

\documentclass[prb,twocolumn,floatfix,preprintnumbers,amsmath,amssymb,superscriptaddress]{revtex4}

\usepackage{graphicx}% Include figure files
\usepackage{dcolumn}% Align table columns on decimal point
\usepackage{bm}% bold math
\usepackage{color}
\usepackage{xspace}

\begin{document}

\title{Criticality in a disordered quantum antiferromagnet by neutron diffraction}

\author{E. Wulf}
\affiliation{Neutron Scattering and Magnetism, Laboratory for Solid State
Physics, ETH Z\"urich, Z\"urich, Switzerland.}

\author{D. H{\"u}vonen}
\affiliation{Neutron Scattering and Magnetism, Laboratory for Solid State
Physics, ETH Z\"urich, Z\"urich, Switzerland.}

\author{J.-W. Kim}
\affiliation{National High Magnetic Field Laboratory, MPA-CMMS group, Los Alamos National Lab (LANL), Los Alamos, NM 87545, USA.}

\affiliation{Lujan Center for Neutron Scattering, LANL, Los Alamos, NM 87545, USA.}

\author{A. Paduan-Filho}
\affiliation{High Magnetic Field Laboratory, University of S\~ao Paulo,
05315-970, S\~ao Paulo, Brazil.}

 \author{E. Ressouche}
 \affiliation{INAC/SPSMS-MDN, CEA/Grenoble, 17 rue des Martyrs, 38054 Grenoble Cedex 9, France.}

\author{S. Gvasaliya}
\affiliation{Neutron Scattering and Magnetism, Laboratory for Solid State
Physics, ETH Z\"urich, Z\"urich, Switzerland.}

\author{V. Zapf}

\affiliation{National High Magnetic Field Laboratory, MPA-CMMS group, Los Alamos National Lab (LANL), Los Alamos, NM 87545, USA.}

\author{A. Zheludev}
 \email{zhelud@ethz.ch}
 \homepage{http://www.neutron.ethz.ch/}
\affiliation{Neutron Scattering and Magnetism, Laboratory for Solid State
Physics, ETH Z\"urich, Z\"urich, Switzerland.}

\date{\today}

\begin{abstract}
Field-induced magnetic ordering in the structurally disordered quantum magnets \DTNX, $x=8$\% and $13$\%, is studied by means of neutron diffraction. The order parameter critical exponent is found to be very close to its value $\beta=0.5$ expected for magnetic Bose-Einstein condensation in the absence of disorder. This result applies to temperatures down to 40~mK, and a 1~T range in magnetic field. The crossover exponent is found to be $\phi \sim 0.4$, for temperature ranges as small as $300$~mK. Possible reasons for a discrepancy with recent numerical simulations and bulk measurements are discussed.
\end{abstract}

 \maketitle

\section{Introduction}

It has been known for decades that disorder may qualitatively alter the universality class of a phase transition, if only certain requirements, such as the Harris criterion, \cite{Harris1974} are satisfied. With a more recent focus on {\it quantum} phase transitions,\cite{Sachdevbook} we now ask how disorder may alter the scaling properties of quantum critical points (QCPs).\cite{Vojta2013} Arguably, the most celebrated QCP is that of Bose-Einstein condensation. In a seminal paper,\cite{Fisher1989} M. Fisher {\it et al.} have made specific predictions for critical exponents of BEC in a random disorder potential, and revealed a novel state of disordered Bosonic matter, the so-called Bose Glass (BG). \cite{Giamarchi1987,Giamarchi1988,Fisher1989} Subsequent theoretical studies and numerical simulations \cite{Priyadarshee2006,Hitchcock2006,Weichman2008,Yu2010,Yu2012} confirmed these predictions, but  also questioned some crucial details. In particular, the  actual values of the critical exponents\cite{Fisher1989,Priyadarshee2006,Weichman2008,Yu2012,Yu2012-2} and even the very existence of the BG phase\cite{Rakhimov2012} are still a matter of hot debate. This makes an experimental study of the BG to BEC QCP  particularly important.

In a parallel effort, quantum simulators of BEC were found among magnetic insulators, particularly in gapped Heisenberg or XY antiferromagnets.\cite{Giamarchi2008} QCPs induced in these real-world materials by an application of external magnetic fields are described in terms of a BEC of elementary magnetic quasiparticles. A recent idea is to introduce chemical disorder into such compounds, to simulate a random potential.\cite{Zheludev2013} Indeed, experiments on several systems including \IPACX (abbreviated IPACX),\cite{Manaka2008,Manaka2009,Hong2010PRBRC} \TCX \cite{Yamada2011,Zheludev2011,Yamada2011reply} and \PHCX (abbreviated PHCX) \cite{Huevonen2012,Huevonen2012-2,Huevonen2013} revealed a field-induced BG-like state, and confirmed a modification of critical exponents. Unfortunately, the presence of small but potentially relevant magnetic anisotropies complicates any direct quantitative comparisons with the axially symmetric case of BEC.\cite{Zheludev2013}

\begin{figure}
\includegraphics[width=\columnwidth]{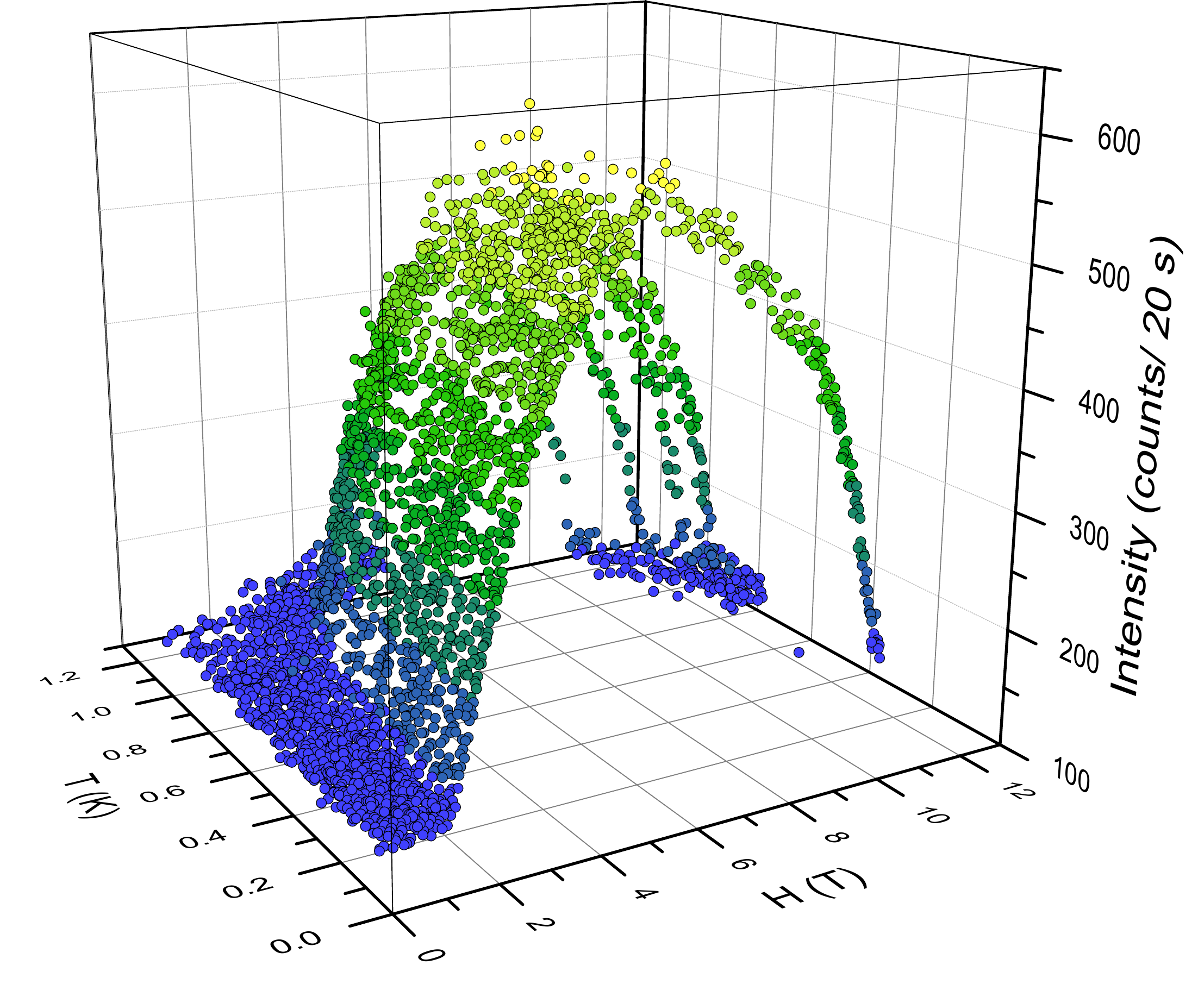}
\caption{(Color online)Neutron intensity measured in the $x=8\%$ \DTNX sample at the $(0.5,0.5,0.5)$ Bragg position, vs. temperature and magnetic field applied along the $c$ axis. \label{Fig:alldata}}
\end{figure}

The anisotropy problem was largely overcome in recent ground-breaking studies of the {\it tetragonal}, and therefore axially symmetric, disordered spin system \DTNX (here abbreviated as DTNX).\cite{Yu2012} An experimental determination of the BEC phase boundary from bulk measurements provided an estimate of the key crossover exponent $\phi\approx 1.1$.\cite{Yu2012} This value is substantially different from $\phi=2/3$ for the magnetic BEC transition in the absence of disorder,\cite{Nikuni2000,Giamarchi2008} but in good agreement with quantum Monte Carlo (QMC) simulations for the disordered system.\cite{Yu2012} In the present study, we determine this phase boundary using a more direct method, namely neutron diffraction which specifically probes the magnetic BEC order parameter. In addition to $\phi$, we also measure the order parameter critical exponent $\beta$.

\section{Materials and experimental considerations}

The tetragonal crystal structure of DTNX and its parent disorder-free compound \DTN are described in detail in Ref.~\onlinecite{Yu2012}. The magnetic properties are due to Ni$^{2+}$ $S=1$ ions. Due to a strong single-ion XY anisotropy, the ground state is a spin singlet, with each Ni spin in the $S_z=0$ state. The lowest energy excitations are due to $S_z=0\rightarrow S_z=\pm1$ transitions. Due to exchange interactions between ions, they are dispersive, with an energy minimum at the antiferromagnetic (AF) zone-center $(0.5,0.5,0.5)$.\cite{Zapf2006,Tsyrulin2013} At this point spectroscopy experiments observe a gap $\Delta=0.29$~meV in the parent material. The BEC transition is induced by an application of an external magnetic field $H$ along the tetragonal axis,\cite{Paduan2004} which at $H>H_c=2.1$~T drives the energy of the $S_z=1$ member of the excitation doublet to zero. The result is spontaneous long-range magnetic order in the $(x,y)$ plane which in neutron diffraction experiments gives rise to a new magnetic Bragg reflection at the AF zone-center.\cite{Tsyrulin2013} The intensity of this peak is proportional to the square of the BEC order parameter. In \DTNX the critical field is reduced compared to the parent material.\cite{Yu2012} In the present work we have performed experiments on fully deuterated single crystal samples, with a nominal Br concentration 13\% (sample A, 218~mg) and 8\% (sample B, 226~mg), respectively. The Br content was confirmed by single crystal X-ray diffraction studies of small fragments detached from the samples.

\begin{figure}
\includegraphics[width=\columnwidth]{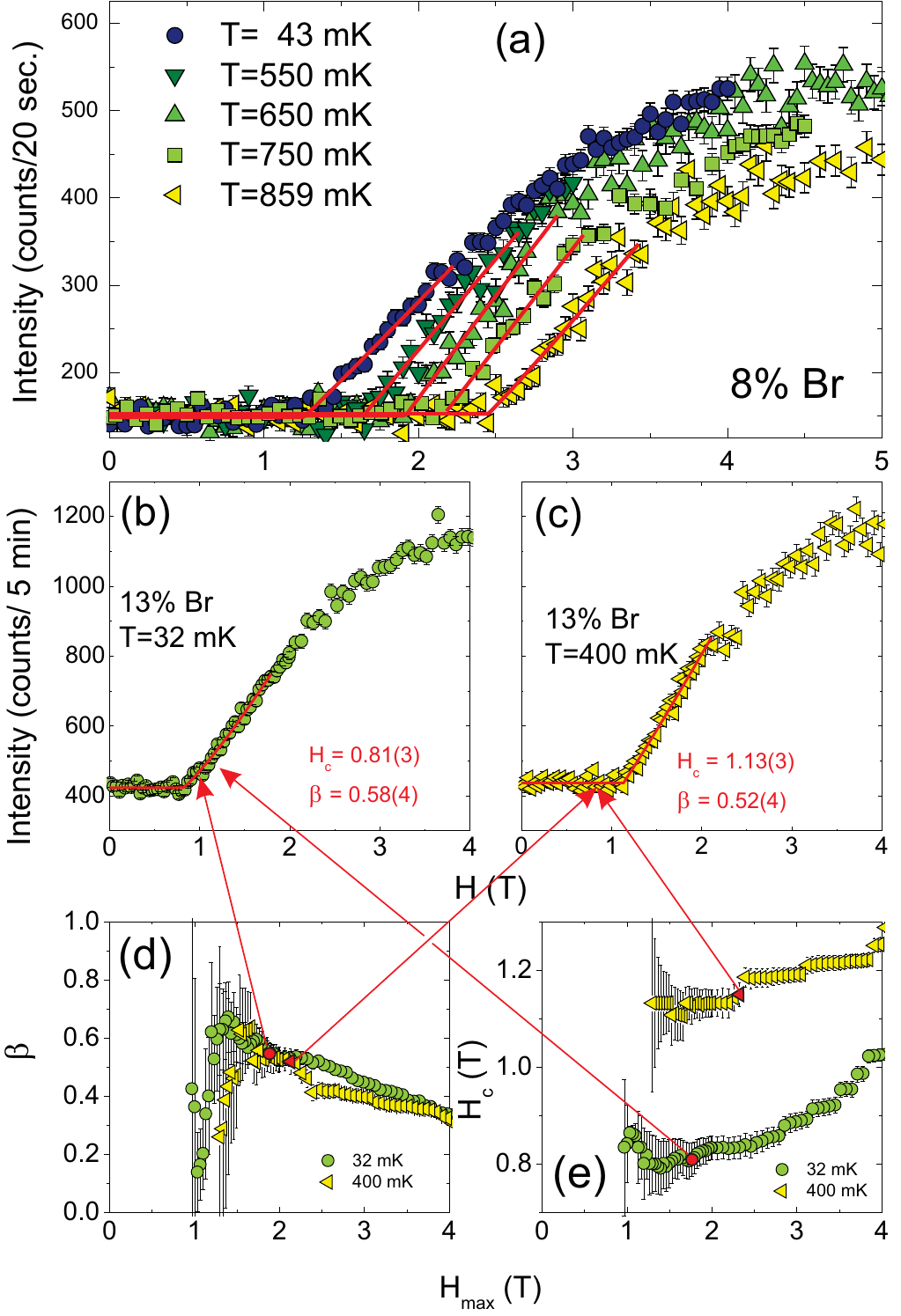}
\caption{(Color online) (a-c) Symbols: typical constant-temperature fields scans of the magnetic Bragg peak intensity in  $x=8\%$ (a) and $x=13$\% (b,c) \DTNX samples. Solid lines in are power-law fits, as described in the text. (d,e) Fitted order parameter critical exponent $\beta$ (d) and critical field $H_c$ (e) for the data in graphs (b) and (c), plotted as a function of maximum field $H_\mathrm{max}$ used in the fitting process. The highlighted symbols illustrate the choice of narrowest useful fitting range, as described in the text. \label{Fig:exdata}}
\end{figure}

The geometry of neutron diffraction experiments require special consideration. First, to reach the BEC QCP, rather than an Ising-type QCP, the magnetic field is to be aligned strictly along the $c$ axis.\cite{Tsyrulin2013} Second, the magnetic propagation vector has a large component along that direction, which is not trivial to reconcile with the geometries of typical superconducting magnets used in neutron experiments.
In our experimental Setup 1, used for Sample A, an appropriate geometry was realized on the TASP 3-axis spectrometer at Paul Scherrer Institut, using the standard horizontal-plane scattering geometry, a Pyrolitic Graphite PG(002) monochromator, $3.1$~meV neutrons, a cooled Be higher order filter, guide-$80'$-$80'$-$80'$ collimation for improved resolution,  and a $7$~T {\it horizontal}-field magnet. The typical experimental wave vector resolution 0.03~\AA$^{-1}$ allows to detect intrinsic peak widths corresponding to correlation length smaller than approximately 200~\AA, or about 25 lattice units, in real space.
The data for Sample B were collected on the D23 diffractometer at Institut Laue Langevin in the {\it lifting counter} geometry, with a 12~T vertical field split-coil cryomagnet, a PG monochromator and 14~meV neutrons.  In all cases sample environment was a $^3$He-$^4$He dilution refrigerator. The alignment of the crystals with respect to the applied field was better than $1^\circ$, as determined {\it in situ} by diffraction measurements.

\section{Results and discussion}

In both samples, magnetic Bragg peaks were observed at low temperatures in sufficiently high magnetic fields at the $(0.5,0.5,0.5)$ position. No differences of intensity were observed between field-cooled (FC) and zero-field-cooled (ZFC) protocols. Moreover, wave vector scans  revealed that the reflections are resolution-limited at all times. Such behavior is in stark contrast with previous experiments on IPACX \cite{Hong2010PRBRC} and PHCX, \cite{Huevonen2012} where the high field phase has history-dependent short range order only. The reproducibility and narrowness of the Bragg reflections in DTNX give us confidence that the high field order phase is indeed a long range ordered magnetic BEC.

\begin{figure}
\includegraphics[width=\columnwidth]{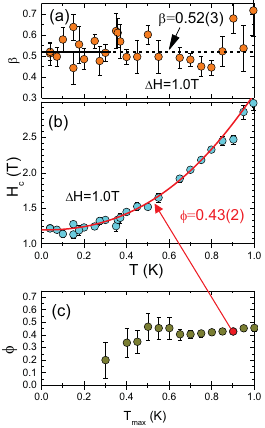}
\caption{(Color online) Data analysis for the $x=8$\% \DTNX sample. (a) Symbols: temperature dependence of the order parameter exponent $\beta$ obtained in fits to individual fit scans in the range $0<H<H_c+\Delta H$, $\Delta H=1.0$~T. The line represents an average obtained for all data measured at $T\lesssim 0.3$~K. (b) Symbols: temperature dependence of the critical field $H_c$ obtained in fits to individual fit scans in the range $0<H<H_c+\Delta H$, $\Delta H=1.0$~T. The line is a power-law fit in the range $0<T<T_\mathrm{max}$, $T_\mathrm{max}=0.85$~K, as described in the text. (c) Fitted value of the critical index $\phi$ as a function of fitting interval. The solid symbol represents the narrowest useful fitting range, as described in the text. \label{Fig:results}}
\end{figure}

The critical indexes were determined from measurements of magnetic Bragg peak intensities in a series of field scans at constant temperatures. The more comprehensive data set for Sample B is shown in Fig.~\ref{Fig:alldata}. Representative scans for samples A and B are plotted in symbols in Fig.~\ref{Fig:exdata}(a-c). In each sample, for each temperature, the transition field $H_c$ and critical exponent $\beta$ were determined in power law least squares fits using the procedure described in Ref.~\onlinecite{Huevonen2012}. For each intensity vs. field scan $I(H)$, the adjustable parameters are the critical field $H_c$, exponent $\beta$, $I(H)\propto (H-H_c)^{2\beta}$, a proportionality coefficient, and a field-independent flat background.

The fits were performed in progressively shrinking fitting intervals defined by the maximum field $H_\mathrm{max}$. Choosing $H_\mathrm{max}$ has to balance the need to stay within the potentially narrow critical region on the one hand, and to obtain the smallest possible error bar on the fitted values on the other hand.
As explained in Ref.~\onlinecite{Huevonen2012}, a reasonable choice is a fitting range, such  that shrinking it further, no longer produces a statistically significant change in the fitted parameters. Selecting such ``minimal useful fitting ranges'' for the data shown in Fig.~\ref{Fig:exdata}(b,c) is illustrated in Fig.~\ref{Fig:exdata}(d,e). The latter show the fitted values of $\beta$ and $H_c$ plotted against $H_\mathrm{max}$, respectively. The highlighted symbols mark the minimal useful fitting ranges for the two data sets. For each data set, any fits in narrower ranges yield results containing these ones within the error bars.
Fitted curves corresponding to the thus selected ranges are shown in red lines in Fig.~\ref{Fig:exdata}(b,c).

For different scans obtained in either sample, the minimal useful fitting ranges varied between 0.7~T and 1~T beyond the fitted $H_c$. To enable a meaningful comparison between different temperatures, we selected a single fittining interval of $0<H<H_c+\Delta H$, $\Delta H= 1$~T, in a final analysis of all data for both samples. The corresponding fits to representative field scans are shown in solid lines in Fig.~\ref{Fig:exdata}a.

The fit results for Sample A, and some typical values for Sample B, are summarized in Table~\ref{Table}. For Sample B, where a more complete temperature coverage was obtained, the temperature evolution of the  critical exponent $\beta$ is shown in Fig.~\ref{Fig:results}(a). The index appears to be $T$-independent: $\beta \approx 0.5$ . It is very similar for the two samples with different Br content, and consistent with observations for the disorder-free compound,\cite{Tsyrulin2013} despite the differences in $H_c$. In this work, we are primarily interested in $\beta$ at the QCP, i.e., in the limit $T\rightarrow 0$. Fortunately, our ability to measure the exponent {\it improves} at low temperatures, thanks to a larger ordered moment and weaker thermal critical scattering. Specifically, for the 8\% \DTNX sample, previous studied of Ref.~\onlinecite{Yu2012} place our lowest measurement temperatures $\approx 40$~mK well into the regime of the BG-BEC transition. Thus, the experimental result $\beta \approx 0.5$ can be considered applicable to the QCP, at least as far as temperature is concerned.

\begin{table}
\centering
\begin{tabular}{r|r|r|r} %|d|d}
\hline\hline
$x$     & T (mK)  & $H_c$ (Tesla)   & $\beta$       \\
\hline
13\% & 34  & $ 0.82\pm 0.03$ & $0.56\pm 0.04$ \\
13\% &200 & $0.90\pm 0.04$  &$0.54\pm 0.05 $ \\
13\% &400 & $1.13\pm 0.03$  &$0.51\pm 0.04$ \\
\hline
8\% &40  & $ 1.22\pm 0.05$ & $0.51\pm 0.05$ \\
8\% &200 & $1.23\pm 0.04$  &$0.48\pm 0.05 $ \\
8\% &450 & $1.52\pm 0.06$  &$0.50\pm 0.08$ \\
\hline\hline
\end{tabular}
\caption{Critical fields and order parameter critical index obtained in the fit interval $\Delta H=1.0$~T for the $x=13\%$ (top three rows) and $x=8\%$ (bottom three rows) \DTNX samples. \label{Table}}
\end{table}

Our experimentally measured value is very similar to the exact result $\beta=1/2$ for BEC QCP in a 3-dimensional disorder-free system.
On the one hand, this is surprising in view of the numerical findings of Ref.~\onlinecite{Yu2012-2}, which predicts $\beta=0.95$ for the BG-BEC transition. From Fig.~\ref{Fig:exdata}d, it is clear that the experimental best fit remains outside the error bar of this theoretical value for fitting ranges at least as narrow as 0.5~T above $H_c$. On the other hand, we note that $\beta\sim 0.5$ has been observed in other disordered systems like IPACX \cite{Hong2010PRBRC} and PHCX,\cite{Huevonen2012} and is consistent with the  limited diffraction data published for \TCX.\cite{Yamada2011reply}
It is important to point out that the discrepancy with theory can not be blamed on a frequent complication in diffraction measurements, namely critical fluctuations. Indeed, in some cases, critical quasielastic scattering just below the transition point is hard to distinguish from true and purely elastic Bragg scattering. In our case,
consistent results between data obtained on a diffractometer (poor energy resolution) and 3-axis spectrometer (tight energy window) suggest that this artefact does not occur. Even if it did, the fitted value of $\beta$ would be {\it larger} than the actual one.

Another important result of this work is the value of the crossover exponent $\phi$, that we determined for the $x=8\%$ Br sample. This critical index defines the $H-T$ phase boundary: $(H-H_c) \propto T^{1/\phi}$. The critical field $H_c$, determined at each temperature as described above,  is plotted against $T$ in open symbols in Fig.~\ref{Fig:results}(b). As for the field dependence of intensities, these data were analyzed using power law fits in a shrinking temperature range. The fitted value of $\phi$ is plotted vs. the maximum temperature $T_\mathrm{max}$ used in the fits in  Fig.~\ref{Fig:results}(c). The highlighted symbol indicates the minimal useful fitting range which was found to be $T_\mathrm{max} =0.9$~K. Further shrinking the fit window does not produce a statistically significant change in the parameter values. In this range, we get $\phi=0.43(2)$, even smaller than $\phi=2/3$ expected for an idealized disorder-free BEC system.\cite{Nikuni2000}
On the other hand, our value is almost exactly equal to that obtained magnetic susceptibility measurements on the parent compound.\cite{Paduan2004} For $T_\mathrm{max} =0.9$~K we get the following estimate for the critical field at $T=0$: $H_c(0)=1.20(2)$~T.

 Our experimental estimate $\phi\sim 0.4$ is considerably smaller than the numerical prediction $\phi=1.06(9)$ for the disordered system and the result of DTNX bulk susceptibility measurements $\phi=1.1(1)$ of Ref.~\onlinecite{Yu2012}. Moreover, the latter values remain well outside our experimental error bars even for narrower fitting ranges, at least down to $T_\mathrm{max}\sim 0.3$~K. For example, for $T_\mathrm{max}=0.4$~K we get $\phi=0.34(11)$. Even so, the most likely cause for the discrepancy with Ref.~\onlinecite{Yu2012} is the temperature range. Indeed, both numerical simulations and the experimental data of that study indicate a crossover to $\phi\sim 1$ only at the lowest temperatures, below $\sim 0.3$~K. Unlike for $\beta$, which in our experiments we can reliably probe down to $\sim 40$~mK, our neutron data for $\phi$ are not sufficient to draw conclusions for this low temperature regime.

\section{Conclusion}

To summarize, neutron diffraction studies of the disordered spin liquid compound \DTNX yield an order parameter critical index $\beta\sim 0.5$ and $\phi\sim 0.4$, for fitting ranges of $\Delta_H=1$~T ($T=40$~mK) and $T_\mathrm{max}=0.9$~K, respectively. There are numerous potential causes for the discrepancy with previous susceptibility measurements (that actually probe a quantity not directly related to the order parameter) and numerical simulations (that study an idealized model of the actual material). Here we shall only re-emphasize the main potential limitation of the present study, namely the significant fitting ranges in magnetic field and temperatures. If another scaling regime emerges still closer to the QCP than $\Delta H=0.5$~T, it is not accessible to neutron diffraction at the present stage. We hope that our findings will contribute to solving the challenging but very important problem of quantum criticality in disordered spin systems, and stimulate further theoretical work, particularly calculations of the order parameter close to the QCP.

\acknowledgements

This work is partially supported by the Swiss National Fund under
project 2-77060-11 and through Project 6 of MANEP. The authors would like to thank T. Roscilde (ENS Lyon) for many stimulating discussions. VSZ and JWK acknowledge the NHMFL and a User Collaboration Grant funded by the U.S. NSF DMR-1157490, the U.S. DOE, and the State of Florida. JWK acknowledges support from the Lujan Center at LANL, which is operated by the DOE's Office of Basic Energy Sciences.

%\bibliography{c:/home/zhelud/bib/azbib}

\end{document}